\documentclass[12pt]{article}

\usepackage[T1]{fontenc}
\usepackage[cp1250]{inputenc}
\usepackage{graphicx}
\usepackage{wrapfig}
\usepackage{amsmath}
\usepackage{ulem}
\usepackage{bm}
\usepackage{amssymb}
\makeatletter
\usepackage{babel}
\makeatother

\newcommand{\nl}{\hfil\break}
\newcommand{\barr}{\begin{eqnarray}}
\newcommand{\earr}{\end{eqnarray}}

\def\bb#1{{\bf #1}}
\def\~{\tilde}

\usepackage{xcolor}

\begin{document}

\title{Quantum Measurement Without Collapse}
\author{Pavel B\'ona\footnote{email: balek@fmph.uniba.sk}\\
\date{\vspace{-8ex}}}

\maketitle\centerline{\it Department of Theoretical Physics,
Comenius University, Bratislava}

\vspace{3ex}

\begin{abstract}
\noindent It is shown that the classical book by von Neumann
proposing dynamics of measured systems with ``reduction (or
collapse) of system's wave packet'' contains also hints how to
avoid this discontinuity in time evolution of the measured system
(hence it is in this point not quite selfconsistent). The
possibility of avoiding that collapse is a consequence of the
observation that any ``human observer'' can observe simultaneously
just mutually compatible ``observables''. In the paper it is shown
how to describe this fact and its consequences. The proposed
interpretation of quantum measurement leads also to trivial
solution of the ``Schr\"odinger Cat Paradox''.
\end{abstract}

\vspace{3ex}

Let us sketch a simple model of any ``measurement'':

We need to have a ``system'', which is the object of our present
interest intended to be ``measured''. The specification of this
system should distinguish it from other ``systems'', i.e. some
criteria of its identification should be used. For repeated trials
of the experiment, which are unavoidable in such a statistical
branch of science as quantum mechanics (QM), we have to be sure
that repeatedly used ``systems'' are equal to each other
(according to the experimenter's criteria). Such a specification
procedure is the decisive part of ``{\bf preparation}'' of the
system in the first part of experiment. The prepared system is
called to exist in a ``{\bf specific state}''. Next the system is
supposed to move in known environment, so that its state can be
considered to be known from the known laws of nature and
calculated from the corresponding known, generally accepted ``{\bf
equations of motion}''.

The remaining parts of the experiment with the prepared systems
are ``{\bf measurement}'' and ``{\bf reading of the results}''.
These parts of experiments in QM will be our main interest in this
paper. We shall describe a scheme of the measuring process with
observation of results in QM later in this paper, starting with a
brief analysis and criticism of some of its essential features
appearing in \cite{vNeumann}, which seem to be almost generally
accepted in physical community.

Before that, we shall sketch the  key aspects of the general
scheme of the mathematical model of QM.\nl

{\bf The formalism of (nonrelativistic) QM}\nl

QM describes `physical systems', i.e. mainly microsystems like
(stable) elementary particles, atoms, molecules, and finite
collections of them. Each system is specified by an ``\bb{algebra
of observables}'' (a C*-algebra), usually realized as a symmetric
algebra of bounded operators on the Hilbert space $\cal H$
ascribed to the system. Each system can be in any time
characterized by its \bb {state}. States are described by
normalized state-vectors $\psi,\phi,... \in \cal H$ (equivalently
by one-dimensional orthogonal projectors ${\bf P}_\psi, {\bf
P}_\phi ...$) or by the so called density matrices, or density
operators, being positive normalized operators $\rho\in
\cal{T(H)}$: $Tr\,\rho = 1$, where $\cal{T(H)}$ is the linear
space of all trace-class operators on $\cal H$. Each state-vector
$\psi\in\cal H$ is in a physically equivalent way represented by
the density matrix $\rho_\psi := {\bf P}_\psi$, so that we can
work with ``states'' and the ``density matrices'' as with
synonyms.

{\bf Observables} in QM represent the physical quantities
characterizing the described \linebreak system, they are
identified with selfadjoint operators on $\cal H$. It is
postulated in QM that all selfadjoint operators (or at least the
bounded ones) represent observables, hence some measurable
quantities. It is usually unspecified which physical (measurable)
meaning is to be assigned to each selfadjoint operator. It is
often useful to work with abstract setting if the set of
observables forms a C*-algebra (which can be always represented as
an algebra of bounded operators in some Hilbert space), cf.
\cite{sak}.

Each selfadjoint operator is characterized by its {\bf spectral
projection measure E}, i.e. a projector-valued function of Borel
sets on real line $\mathbb{R}$ with values in orthogonal
projectors:

\barr\nonumber E: B\ (\in {\cal{B}}(\mathbb{R})) \mapsto E(B),\ \
E(\mathbb{R})=I_{\cal H};\\ B_j \cap B_k = \emptyset\Rightarrow
E(B_j)E(B_k) = 0, E(\cup_k B_k)=\sum_k E(B_k).\nonumber\earr

If an observable (i.e. a selfadjoint operator) $\frak A$ has
purely discrete spectrum, i.e. there is a complete (i.e. such that
it cannot be extended) orthonormal system of ``{\bf
eigenfunctions}'' or ``{\bf eigenvectors}'' in $\cal H$ $\supset$
$\{\psi_n, n=1,2,...\}:\ (\psi_k|\psi_n) = \delta_{k,n}$, and
$\frak A\,\psi_n = \lambda_n\,\psi_n, \forall n$, then the system
can attain only the values $\lambda_n$ as possible values of the
quantity $\frak A$, i.e. by measuring $\frak A$ only the results
$\{ \lambda_n, n= 1,2,...\}$ can be obtained. If another
observable $\frak B$ can have simultaneosly with $\frak A$ (i.e.
in the same complete set of states) sharp values $\{\mu_n,
n=1,2,...\}$, the operators mutually commute; and vice versa: If
$\frak A \frak B = \frak B \frak A$, then these operators have a
common complete orthonormal set of eigenvectors $\{ \chi_n, n=
1,2,...\}, \frak A \chi_n=\lambda_n\chi_n; \frak B \chi_n=\mu_n
\chi_n.$

Each vector $\varphi\in\cal H$ can be expressed as a linear
combination of any such orthonormal system: $\varphi=\sum_n
c_n\psi_n$, where the coefficients $c_n=(\psi_n|\varphi)$
represent the probabilities $|c_n|^2$ of obtaining the result
$\lambda_n$ by measurement of the observable $\frak A$ in the
state $\varphi$.

QM is an ``{\bf intrinsically statistical theory}'', which means
that there is no state of any system in which all the observables
have specific values: the state-vector or density matrix of a
specific state contains just information on ``{\bf probability
distributions}'' of all observables in that state; a sharp value
of some observable in a given state is a rather exceptional
possibility. This leads to serious problems of understanding and
description of the measurement process: Each time after
measurement of some quantity of a single system (i.e. after just a
single ``{\bf particle}'' was interacting with the measuring
apparatus), the apparatus finds one specific value also in the
case if the measured quantity of the considered system in the
measured state has just probability distribution of possible
values of the quantity with nonzero dispersion. If the observer
wants to obtain the form of that probability distribution by the
measurement, he has to repeat the same measurement on many copies
of ``{\bf equally prepared}'' system under consideration
($\approx$ particle). This leads to questions we intend to discuss
later in this paper.

{\bf The dynamics of QM-systems} (similarly as in classical
mechanics) is specified by the specification of the operator $\bf
H$, the ``{\bf system's Hamiltonian}'' corresponding to the ``{\bf
energy}''. If the considered  system does not interact with other
systems, then the time evolution of its state-vectors is:
$$ \psi(t) = e^{-it\bf H}\,\psi(0),\ \text{or }\ {\bf
P}_{\psi(t)}= e^{-it\bf H}{\bf P}_{\psi(0)}e^{it\bf H}. $$ If the
considered system is not isolated, but interacts with another
system, then this pair of systems is considered as a new larger
isolated system with its own new Hamiltonian, consisting of the
sum of the ``free Hamiltonians'' of each system and, in addition,
of some "interaction Hamiltonian" characterizing the mutual
interaction of the two systems.

The only problematic issue in this abstract general scheme is the
question of some satisfactory description of the evolution of the
considered system in the time of the physical act of measurement
of some specific observable: Besides the ``measured system'' also
some macroscopic apparatus is acting (and interacting), changing
its state so that the {\bf human observer} can read the result of
the measurement. The proposed description of this process leads
also to an answer to the question ``how can the Schr\"odinger Cat
be simultaneously dead and alive''.

\nl {\bf The essential points of the von Neumann scheme of
measurement in QM}\nl

(i)\ The measured system {\bf S}\ in the state $\rho$ interacts
with the measuring macro\-scopic apparatus {\bf M}\ chosen for
measurement of a specific observable $\frak A$. In general,
according to von Neumann, a chain of other similarly working
apparatuses ${\bf M_1+M_2+...}$, all of them described as
quantum-mechanical systems, can be added to {\bf M}, ending either
by a human observer (or even by his brain), or an apparatus in
which the result can be fixed by some ``pointer position'' (and
prepared for observation by human observers).\nl

(ii)\ The states of above mentioned chain of apparatuses are
changed due to interaction with {\bf S}\ and the following
evolution by continuous unitary evolution operator of the composed
system (=process 2, cf. \cite[Ch. V.1 p.230]{vNeumann}). Hence the
measured state $\rho$ is transformed subsequently isomorphically
in the instant of the interaction-measurement by the measuring
device {\bf M} to the states of the measuring systems ${\bf
M+M_1+M_2+...}$ by unitary time evolution. So, we shall call the
subsequently arising states of the apparatuses ${\bf M+M_1+}$
${\bf M_2+...}$ ``the measured state $\rho$ of {\bf S}''. It can
be seen that the whole chain of apparatuses can be considered
effectively as a single apparatus {\bf M}.\nl

(iii)\ The last element of the chain modifies the observed state
$\rho$ by an abrupt change depending on the measured observable
$\frak A$. We shall assume, for simplicity, that $\frak A$ has
purely discrete nondegenerate spectrum with (complete orthonormal)
set of eigenvectors $\{\psi_n, n=1,2,...\}$: $\frak A\,\psi_n =
\lambda_n\,\psi_n, \forall n$. The mentioned abrupt change looks
as follows \cite[p.230]{vNeumann}:

$$\rho \mapsto \rho' := \sum_n (\psi_n|\rho|\psi_n)\,{\bf
P}_{\psi_n}. $$

\noindent This is description of the famous ``{\bf collapse of the
wave packet}''; the change $\rho \mapsto \rho'$ is considered to
be instantaneous. This change cannot be reached by any unitary
transformation, hence it is not accessible by Schr\"odinger-like
continuous evolution with any Hamiltonian (in nontrivial cases
$\rho' \neq \rho$).\nl

(iv)\ The von Neumann's book \cite{vNeumann} also contains,
however, certain notes which could (and will!) enable us to {\it
avoid the above mentioned collapse}. Namely from the
considerations in \cite[p.262]{vNeumann} it clearly follows that
{\bf any final human observer (as the last element of the chain of
{\bf M}'s) can observe simultaneously measurable, hence mutually
commuting observables only.}\footnote{This assertion was even more
explicitly expressed in the original German version of
\cite{vNeumann}, in the paragraph situated between the last two
paragraphs of \cite[p.261]{vNeumann}, which was omitted in the
English edition.}\nl

\nl {\bf  A proposal for alternative description of quantum
measurement process}\nl

Let us note first that there is an ambiguity in von Neumann's
interpretation of the reduced density matrix: Since the set of
density matrices does not form a {\bf simplex}, the convex
decomposition of nontrivial density matrices to extremal elements
is highly nonunique. Moreover, various decompositions of a given
density matrix may be incompatible: Different decompositions may
contain states corresponding to eigenvectors of mutually
noncommuting observables. If it is possible to describe the
presently considered physical situation by elements of a simplex
which are unambiguously decomposable into extremal elements having
clear physical meaning, then these elements can be interpreted as
``{\bf proper mixtures}''. We shall see that the state described
by the reduced density matrix can be naturally transformed into
such a form.

Accepting (essentially) the point (ii) of our description of the
von Neumann scheme of the measurement on a system {\bf S}\
occurring in a state $\psi:= \sum_j c_j\psi_j$ by a measuring
(quantum macroscopic) apparatus {\bf M}, we assume that the
interaction of {\bf S+M}\ leads to such a change of the state of
the combined system where the macroscopically observable and
mutually distinguish\-able state-vectors $\Phi_j$ corresponding to
the values $\lambda_j$ of the observable $\frak A$ and to the
states of $\psi_j$ of {\bf S} (i.e. representing the corresponding
``pointer positions'') are (via unitary mapping by U) in one to
one correspondence with the states $\psi_j$. This is in agreement
with \cite[p.285-286]{vNeumann}:

$$ U:\ \psi\otimes\Phi_0 = \sum_j c_j\psi_j\otimes \Phi_0 \mapsto
\sum_j c_j\psi_j\otimes \Phi_j, $$

\noindent where $\Phi_0$ is the state of {\bf M} just before the
measurement interaction and $\Phi:=\sum_j c_j\Phi_j$ is the state
of {\bf M} immediately after (or in the instant of, if the
measurement is instantaneous) the measurement, obtained by partial
trace of the above described ``two-systems state-vector'' w.r.t.
the measured system, cf. \cite{davies}.

The following fate (i.e the subsequent dynamical evolution) of the
state of the system {\bf S} after the interaction with macrosystem
{\bf M} might be very diverse, c.f. \cite[\S 7, esp.
p.24]{LandQM}: {\bf S} could be absorbed by the macrosystem {\bf
M}, or just dispersed by {\bf M} with various simultaneous changes
of its state. For the result of the measurement this is not
important. However, such a simple (and even general) scheme that
the measured state of {\bf S} after its interaction with the
macroscopic measuring device {\bf M} will continue its dynamical
evolution, just ``reduced'' to the eigenstate of the measured
quantity corresponding to the obtained measured eigenvalue, is
rather popular but nonrealistic simplification and deformation of
the process.

What is important, however, is the imprint left by {\bf S} on {\bf
M}: Because the resulting ``pointer positions'' after each single
run of the experiment should be repeatedly visible by different
people, the states $\Phi_j$ should be `sufficiently stable', or
stationary w.r.t. the dynamics (of at least macroscopic
observables) of {\bf M}. Then the observable of {\bf M} with
eigenvectors $\Phi_j (j= 1,2,\dots)$ can be denoted by $\~{\frak
A}$, because it is `a copy` of the corresponding observable $\frak
A$ of the measured system {\bf S}. So, this $\~{\frak A}$, or its
eigenvalues, are what has to be observed by any human observer
taking part in the considered measuring process. But according to
(iv) above, {\bf all the observables of M that are simultaneously
observable by any human observer must, as the corresponding
operators, mutually commute}.\nl

Let us form from these mutually observable quantities of {\bf M},
necessarily including our macroscopic $\~{\frak A}$, a C*-algebra
$\cal A$. Here there is some ambiguity in our construction -- the
choice of the C*-algebra might depend on the ability of observers
to observe  certain observables from $\cal A$.\footnote{This
corresponds to the ambiguity of the definition of {\bf macroscopic
observables} in finite systems; they cannot be unambiguously
defined in an abstract theory.} We could simply choose $\cal A$ as
the maximal commutative subalgebra of all observables of {\bf M}
containing $\~{\frak A}$; but this choice would necessarily
contain also microscopic variables of the macroscopic system {\bf
M}. So, let us take some `convenient' abelian C*-algebra $\cal A$
containing $\~{\frak A}$, e.g. the C*-algebra with unit element
generated by $\~{\frak A}$ (as the minimal possibility).

What are the states of {\bf M} accessible to our human observers?
These are the states, i.e. positive normalized linear functionals,
of the abelian C*-algebra $\cal A$. Let us describe the state
space of $\cal A$, cf. \cite{{sak}, {najm}}: For each abelian
C*-algebra $\cal A$ there exists its {\bf spectrum space} $\cal
M$, i.e. a compact subspace in the w*-topology of the whole state
space $\cal S(A)$ of $\cal A$, consisting of its extremal (pure)
states. The points of $\cal M$ are in bijective correspondence
with the pure (=extremal) states on $\cal A$. The states $\cal
S(A)$ consist of all probability Radon measures on $\cal M$, which
form a (Choquet) simplex (in the same sense as classical
probability distributions on the phase space of classical
mechanics form a simplex: each such probability distribution is a
convex combination-integral of the functions supported by single
points ($\delta$ functions)). This is valid for an arbitrary
choice of the commutative C*-algebra. Note that the characteristic
property of any simplex is that it is a subset of a linear space
such that each of its elements has {\bf unequivocal} convex
decomposition into its extremal elements.\nl

Let us see how the state ${\bf P}_\psi$ of {\bf S} corresponding
to the measured state-vector $\psi := \sum_jc_j\psi_j$, or its
unitary ``macroscopic copy'' $\Phi$, as described above, looks
from the point of view of a human observer. In the basis $\Phi_j
(j= 1,2,\dots)$ -- the unitary ``copy'' of $\psi_j (j= 1,2,\dots)$
-- the measured state-vector $\psi = \sum_j c_j \psi_j$ is
transformed into:

$$ \Phi := \sum_j c_j \Phi_j. $$

\noindent Since we are interested in the states on $\cal A$ only,
we have to restrict the state ${\bf P}_\Phi$ as a linear
functional on all the observables of {\bf M} to the observables
from $\cal A$. The transformed macroscopic form $\~{\frak A}$ of
$\frak A$ has in the Hilbert space $\cal H_{\bf M}$ of {\bf M} the
complete orthonormal set of eigenvectors $\Phi_j$. The
``measured'' state $\Phi$ applied to any $a\in\cal A$ (as a linear
functional on $\cal A$) yields

$$ \Phi: a\mapsto Tr(P_\Phi a) = (\Phi| a |\Phi) = (\sum_j c_j \Phi_j| a
|\sum_k c_k\Phi_k) = \sum_j |c_j|^2 (\Phi_j| a |\Phi_j), \forall
a\in\cal A,
$$

\noindent because $\{ \Phi_j: j=1,2,\dots\}$ is an orthonormal set
of eigenvectors of all the elements of $\cal A$.

We see that a ``pure state-vector'' on the whole algebra of
observables $L(H_{\bf M})$ decomposes into a genuine (=proper)
mixture of the states corresponding to single elements of the
``eigenbasis'' of the commutative algebra $\cal A$. This is
identical to the above described von Neumann's ``collapse of the
wave packet'' for the case $\rho := {\bf P}_\Phi$.\nl

{\bf Is the Schr\"odinger Cat simultaneously dead and alive ?}\nl

Accepting that any human observer of macroscopic (hence visible)
system can distinguish by his senses mutually commuting
observables only, we have also trivial resolution of the ``Problem
of the Cat''.

Let us have two macroscopic state-vectors $\Psi_1,\ \Psi_2$ of the
observed system, which are mutually orthogonal (e.g. `dead and
alive cat', or two orthogonal states of a very long spin chain).
For any observable $\frak A$ for which these states are
eigenstates it holds

$$ (\Psi_1|\frak A|\Psi_2) = 0. $$

\noindent This is valid for all $\frak A$'s from any commutative
algebra. For a state of the form

$$ \Psi := c_1\Psi_1 + c_2\Psi_2,\ c_1c_2\neq 0 $$

\noindent we have

$$ (\Psi |\frak A|\Psi)= |c_1|^2 (\Psi_1|\frak A|\Psi_1)+ |c_2|^2
(\Psi_2|\frak A|\Psi_2), $$

\noindent which describes a mixture applied to any given element
of the commutative algebra of ``observable observables''. In the
case of the Cat this expresses the situation of looking at the Cat
(statistically, in the sense of QM) and observing the two
possibilities: the Cat is either dead or alive, with probabilities
$|c_1|^2$ and $|c_2|^2$ respectively, but not both simultaneously.
The above expression corresponds to a genuine (=proper) mixture of
the states $\Psi_1, \Psi_2$.

This also shows how to resolve the problem of possible
interferences of states of large but finite subsystems of an
infinite system, e.g. a spin chain. We can consider, at least
``ideologically'', problems like those arising e.g. in \cite[Sec.
7.6]{bon-book} in the construction of ``Models of Quantum
Measurement'' to be solved.

\end{document}